\documentclass[runningheads]{llncs}
\usepackage[T1]{fontenc}
\usepackage{graphicx}
\usepackage{cite}
\usepackage{multirow}
\usepackage{amssymb}
\usepackage{amsmath}
\usepackage[pagebackref,breaklinks,colorlinks]{hyperref}
\usepackage[table]{xcolor}
\usepackage{booktabs}

\usepackage[capitalize]{cleveref}
\crefname{section}{Sec.}{Secs.}
\Crefname{section}{Section}{Sections}
\Crefname{table}{Table}{Tables}
\crefname{table}{Tab.}{Tabs.}

\begin{document}
\title{DWA: Differential Wavelet Amplifier for Image Super-Resolution}
%
%
\author{Brian B. Moser\inst{1, 2} \and
Stanislav Frolov\inst{1,2} \and
Federico Raue\inst{1} \and
Sebastian Palacio\inst{1} \and
Andreas Dengel\inst{1, 2}}
\authorrunning{Moser et al.}
%
\institute{German Research Center for Artificial Intelligence (DFKI), Germany \and
RPTU Kaiserslautern-Landau, Germany \\
\email{first.second@dfki.de}}
\maketitle              
\begin{abstract}
This work introduces Differential Wavelet Amplifier (DWA), a drop-in module for wavelet-based image Super-Resolution (SR). 
DWA invigorates an approach recently receiving less attention, namely Discrete Wavelet Transformation (DWT).
DWT enables an efficient image representation for SR and reduces the spatial area of its input by a factor of 4, the overall model size, and computation cost, framing it as an attractive approach for sustainable ML.
Our proposed DWA model improves wavelet-based SR models by leveraging the difference between two convolutional filters to refine relevant feature extraction in the wavelet domain, emphasizing local contrasts and suppressing common noise in the input signals. 
We show its effectiveness by integrating it into existing SR models, e.g., DWSR and MWCNN, and demonstrate a clear improvement in classical SR tasks.
Moreover, DWA enables a direct application of DWSR and MWCNN to input image space, reducing the DWT representation channel-wise since it omits traditional DWT.

\end{abstract}
\section{Introduction}
Image Super-Resolution (SR) has an impressive legacy in Computer Vision (CV) yet still presents an exhilarating challenge \cite{9044873, 10041995}. 
SR is a task of enhancing Low-Resolution (LR) images to High Resolution (HR). 
It is challenging because many High Resolution (HR) images can correspond to a given Low-Resolution (LR) image, rendering the task mathematically ill-posed.

In recent years, deep learning has fueled rapid development in SR, leading to tremendous progress \cite{dong2015image, dong2016accelerating}. 
While many techniques have improved the overall quality of image reconstructions, there remains a pressing need for methods capable of producing high-frequency details, particularly when dealing with high magnification ratios \cite{saharia2022image}. 
Addressing this issue is crucial for the continued advancement of SR.
Influenced by achievements on other CV tasks, recent research focused on trending approaches like Transformer-based networks \cite{vaswani2017attention, liang2021swinir, sun2023spatially}, Denoising Diffusion Probabilistic Models \cite{li2022srdiff, saharia2022image, sahak2023denoising} or Generative Adversarial Networks \cite{wang2018esrgan, Wang_2021_ICCV}.
Despite astonishing reconstruction capabilities, they often lack an explicit focus on generating high-frequency details, i.e., local variations.

This work aims to advance the field of SR by exploring wavelet-based networks. Unfortunately, this technique has received less attention despite its significant potential \cite{10041995}. 
We seek to provide a fresh perspective and revive research by re-evaluating these approaches.
Discrete Wavelet Transformation (DWT) enables an efficient image representation without losing information compared to its na\"ive spatial representation, i.e., traditional RGB format.
It does so by separating high-frequency details in distinct channels and reducing the spatial area of input image representation by a factor of 4.
Therefore, a smaller receptive field is required to capture the input during feature extraction.
Using DWT like in DWSR \cite{guo2017deep} and MWCNN \cite{liu2018multi} reduces the overall model size and computational costs while performing similarly to state-of-the-art image SR architectures.

This work introduces a new Differential Wavelet Amplifier (DWA) module inspired by differential amplifiers from electrical engineering \cite{agarwal2005foundations}. 
Differential amplifiers increase the difference between two input signals and suppress the common voltage shared by the two inputs, called Common Mode Rejection (CMR) \cite{horowitz1989art}. 
In other words, it mitigates the impact of noise (e.g., electromagnetic interference, vibrations, or thermal noise) affecting both source inputs while retaining valuable information and improving the integrity of the measured input signal.
Our proposed DWA layer adapts this idea to deep learning and can be used as a drop-in module to existing SR models. 
This work shows its effectiveness as exemplary for wavelet-based SR approaches.
DWA leverages the difference between two convolutional filters with a stride difference to enhance relevant feature extraction in the wavelet domain, emphasizing local contrasts and suppressing common noise in the input signals.
We demonstrate the effectiveness of DWA through extensive experiments and evaluations, showing improved performance compared to existing wavelet-based SR models without DWA:
DWSR with DWA shows overall better performance w.r.t. PSNR and SSIM, and MWCNN with DWA achieves better SSIM scores with comparable PSNR values on the testing datasets Set5 \cite{bevilacqua2012low}, Set14 \cite{zeyde2010single}, and BSDS100 \cite{martin2001database}.

Taken together, our work makes the following key contributions:
\begin{itemize}
 \item Introduction of Differential Wavelet Amplifier (DWA): a novel module that leverages the difference between two convolutional filters horizontally and vertically in a wavelet-based image representation, which is applicable as drop-in addition in existing network architectures. 
 \item Comprehensive evaluation demonstrating the improved performance by using DWA on popular SR datasets such as Set5 \cite{bevilacqua2012low}, Set14 \cite{zeyde2010single}, and BSDS100 \cite{martin2001database} by adding DWA to existing wavelet-based SR models, namely, DWSR \cite{guo2017deep} and MWCNN \cite{liu2018multi}.
 \item Experimental analysis showing that DWA enables a direct application of DWSR and MWCNN to the input space by avoiding the DWT on the input image. 
This application reduces the input channel-wise to 3 instead of 12 channels for RGB images while keeping the spatial reduction benefit of DWT.
\item Visual examination of reconstructions showcasing that the DWSR with the DWA module captures better distinct edges and finer details, which are also closer to the ground truth residuals.
\end{itemize}
\section{Background}
This chapter provides comprehensive background information on 2D Discrete Wavelet Transform (2D-DWT), how SR models (DWSR \cite{guo2017deep} and MWCNN \cite{liu2018multi}) use it, and related work to Differential Wavelet Amplifiers (DWA). 
Additionally, we introduce differential amplifiers from electrical engineering, which inspired our proposed method DWA.

\subsection{Discrete Wavelet Transform in SR}

The 2D Discrete Wavelet Transform (2D-DWT) decomposes an image into four unique sub-bands with distinct frequency components: a low-frequency approximation sub-band and three high-frequency detail sub-bands representing horizontal, vertical, and diagonal details.
Let $x \left[ n \right] \in \mathbb{R}^N$ be a signal. 
The 1D Discrete Wavelet Transformation (1D-DWT) with Haar wavelet passes the input signal first through a half-band high-filter $h \left[ n \right]$ and a low-pass filter $l \left[ n \right]$. 
%
Next, half of the sample is eliminated according to the Nyquist rule \cite{guo2017deep}.
The wavelet coefficients are calculated by repeating the decomposition to each output coefficient iteratively \cite{stephane1999wavelet}. 
In the case of images, it applies $h \left[ n \right]$ and $l \left[ n \right]$ in different combinations, resulting in four function applications.

The DWSR \cite{guo2017deep} SR model exploits the wavelet domain and gets the DWT representation of the interpolated LR image as input.
DWSR is composed of 10 convolution layers that are applied sequentially.
It adds the interpolated LR input as residual for the final reconstruction step, which results in learning only the sparse residual information between the LR and HR domains.

MWCNN \cite{liu2018multi} exploits multi-level DWT (multiple applications of DWT) and utilizes a U-Net architecture \cite{ronneberger2015u}.
DWT replaces all downsizing steps, and the inverse operation of DWT replaces all upsampling steps.
Ultimately, it uses the interpolated LR image as a residual connection for the final prediction.
The standard MWCNN setup consists of 24 convolution layers.

One caveat of DWSR and MWCNN in learning the residual is that they must translate its rich information input to sparse representation, e.g., the average band.
To ease the burden, we present a Differential Wavelet Amplifier, which directly transforms the input into sparse representations inspired by differential amplifiers introduced next.

\subsection{Differential Amplifier}

An electronic amplifier is a standard electrical engineering device to increase a signal's power \cite{agarwal2005foundations}. 
One type of electronic amplifier is the differential amplifier that increases the difference between two input signals and suppresses the common voltage shared by the two inputs \cite{laplante2018comprehensive}.
Given two inputs $V^{-}_{in}, V^{+}_{in} \in \mathbb{R}^N$ and the differential gain of the amplifier $A_d  \in \mathbb{R}$, the output $V_{out}$ is calculated as 
\begin{equation}
    \label{eq:DA}
     V_{out} = A_d \left( V^{+}_{in} - V^{-}_{in}\right) 
\end{equation}

The purpose of differential amplifiers is to suppress common signals or noise sources that are present in multiple input channels while retaining valuable information. 
In the literature, this is called Common Mode Rejection (CMR) and is a critical property in many electrical engineering applications, particularly in systems that measure small signals in the presence of noise or interference, e.g., electromagnetic interference or thermal noise \cite{horowitz1989art}.
Hence, using CMR improves the signal-to-noise ratio, overall system performance, and signal integrity since the system can focus on the relevant differential signals.

\subsection{Differential Convolutions}
Closest to our work is Sar{\i}g{\"u}l et al. \cite{sarigul2019differential}, which applies differential convolutions, i.e., the difference of two convolution layers, to emphasize contrasts for image classification, which is inherently different to image generation tasks such as image SR. 
Despite this, they do not consider a stride difference vital for capturing variations.
%
Knutsson et al. \cite{knutsson1993normalized} theoretically examine a normalized version of differential convolutions also with no stride difference.
Due to the time of publication, they did not try it in the case of deep learning-based image SR.
Newer applications like Canh et al. \cite{canh2019difference} consider learnable parameters to turn the Difference of Gaussians (DoG) \cite{lowe2004distinctive} into a learnable framework, but has the same caveat:
As Knutsson concluded, their approaches can be interpreted as a standard convolution weighted with the local energy minus the ``mean'' operator acting on the ``mean'' data, i.e., a more elaborate convolution operation. 
%

A similarity could also be seen in the approach of residual connections of ResNets \cite{he2016deep} when the kernel parameters have a negative sign.
However, residual connections are different since they force a convolution layer to learn to extract the sparse details that are not apparent in the input. 
In contrast, our proposed method with Differential Wavelet Amplifier (DWA) explicitly produces sparse details by design due to the subtraction operator.
Therefore, DWA does not have to learn what input information should be removed for the residual information. 
It can focus on relevant features that persist when the stride convolution does not detect the same feature, thereby emphasizing local contrast.
%
%
\begin{figure}
    \begin{center}
        \includegraphics[width=.68\textwidth]{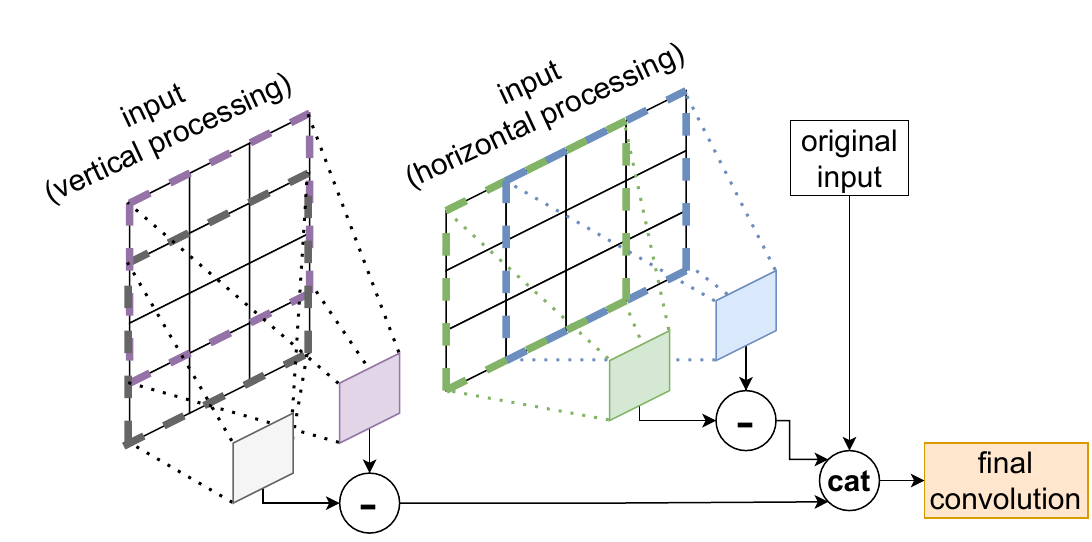}
        \caption{\label{fig:methodology}
        Visualization of DWA. 
        It takes the difference of two convolutional filters with a stride difference of at least 1, vertically and horizontally. Next, it concatenates the input with horizontal and vertical feature maps. In the end, it applies a final convolution.}
    \end{center}
\end{figure}
\section{Differential Wavelet Amplifier (DWA)}


This section presents our proposed Differential Wavelet Amplifier (DWA) module.
Inspired by differential amplifiers in electrical engineering, DWA is designed to operate in the wavelet domain and exploits the difference between two input signals to improve the performance of image SR methods based on wavelet predictions.
%
%
DWA is applied separately in the horizontal and vertical axis of the input image. 
In each direction, we perform two convolutions with a stride distance in one direction for both axis (from left to right, from top to bottom, as in MDLSTMs \cite{graves2008offline}), allowing a fine-grained feature extraction and emphasizing local contrasts while suppressing the common mode in the input, similar to CMR in electrical engineering.
\autoref{fig:methodology} visualizes all processes involved in DWA.

Let $\mathbf{x} \in \mathbb{R}^{w \times h \times c_{in}}$ be an input image or feature map with $c_{in}$ channels. 
We define $\psi \left(\mathbf{x}, (i, j) \right) : \mathbb{R}^{w \times h \times c_{in}} \times \mathbb{N}^2 \rightarrow \mathbb{R}^{k \cdot k \times c_{in}}$ as a function that extracts $k \cdot k$ points around a spatial position $(i, j)$.
We can then express the resulting feature maps for the horizontal $\mathbf{H} \left( \mathbf{x} \right)$ and vertical $\mathbf{V} \left( \mathbf{x} \right)$ axis as 
\begin{equation}
\begin{split}
    \mathbf{H} \left( \mathbf{x} \right)_{i,j} &= f \left( \psi \left(\mathbf{x}, (i, j) \right) ; \theta_1 \right) - f \left( \psi \left(\mathbf{x}, (i+s, j) \right) ; \theta_2 \right), \\
    \mathbf{V} \left( \mathbf{x} \right)_{i,j} &= f \left( \psi \left(\mathbf{x}, (i, j) \right) ; \theta_3 \right) - f \left( \psi \left(\mathbf{x}, (i, j+s) \right) ; \theta_4\right), 
\end{split}
\label{eq:dwa}
\end{equation}
where $f : \mathbb{R}^{k \cdot k \times c_{in}} \rightarrow \mathbb{R}^{c_{f}}$ is a convolution operation with parameters $\theta_n$ for $0 < n < 4$ , $k \times k$ the kernel size and $s \in \mathbb{N}$ a pre-defined stride difference. 
As a result, the local variance is captured in one direction for both axes, similar to MDLSTMs \cite{graves2008offline}: from left to right with parameters $\theta_1$ and $\theta_2$ and from top to bottom with parameters $\theta_3$ and $\theta_4$.
%
We obtain two distinct feature maps that capture complementary input image information and provide richer feature representations for the wavelet-based SR task. 
The input is directly translated to sparse representations, which reduces the distance to residual target objectives in networks that use residual connections for final prediction.

We concatenate the resulting feature maps alongside the input to ensure no information is lost during the DWA processing. 
This combination creates a comprehensive set of feature maps that retains the original input information while incorporating the directional features obtained from both axes. 
More formally:
\begin{equation}
    g \left( \mathbf{x} \right) = \mathbf{x} \odot \sigma \left( H \left( \mathbf{x} \right) \odot V \left( \mathbf{x} \right) \right),
\end{equation}
where $\odot$ is a channel-wise concatenation operator and $\sigma$ is a non-linear function like sigmoid, tanh or ReLU \cite{agarap2018deep}.
%
%
The concatenated feature map is fed into an additional convolution layer $f_{final}: \mathbb{R}^{k \cdot k \times (c_{in} + 2 \cdot c_{f})} \rightarrow \mathbb{R}^{c_{final}}$ and parameters $\theta_{final}$, which maps the channel size after concatenation to a desired target channel size $c_{final}$ such that our module can easily be incorporated into existing models:

\begin{equation}
    \text{DWA} \left( \mathbf{x} \right)_{i,j} = f_{final} \left( \psi \left(g \left(\mathbf{x} \right), (i, j) \right) ; \theta_{final}\right)
\end{equation}

A SR model utilizing this DWA module exploits the comprehensive feature map to learn the complex relationships between LR and HR images, ultimately reconstructing the HR image with reduced noise.
By employing the DWA, we aim to harness the benefits of wavelet domain processing and the difference between two convolutional filters.
We demonstrate the effectiveness of our approach through extensive experiments and evaluations in the following sections.

\subsection{Direct Application of DWA (DWA Direct)}
\label{sec:direct}
One way to circumvent additional computation steps is to apply DWA directly on the image space, omitting DWT and learning the transition between image and frequency space implicitly via DWA.
Thus, the interpolation of the input, which effectively adds no additional information since it generates only approximated values, can be reduced by half for networks like DWSR or MWCNN.
Consequently, the network is better adapted to the given values of the LR input.
In the experiments, we evaluate this alternative approach called DWA Direct and show that it further enhances the performances of DWSR and MWCNN.

\section{Experiments}
We evaluate our proposed DWA module by integrating it into the wavelet-based SR models DWSR and MWCNN.
We begin this section by describing the experiments.
Next, we discuss the results quantitatively and qualitatively.
We show the effectiveness of DWA and that a direct application of wavelet-based SR models with DWA to image space is feasible without forfeiting reconstruction quality.

\subsection{Experimental Setup}
We applied widely-used SR datasets to evaluate our method. In addition, we utilized standard augmentation techniques such as rotation, horizontal and vertical flipping. 
For testing, we employed the datasets Set5 \cite{bevilacqua2012low}, Set14 \cite{zeyde2010single}, BSDS100 \cite{martin2001database}.
For training, we used different settings for DWSR and MWCNN to match the original works for a fair comparison, as dissected in the following.
In all experiments, we train using the Adam optimizer \cite{kingma2014adam} with a learning rate of $10^{-4}$ with $L2$ regularization of $10^{-8}$ on a single A100 GPU.
Moreover, we use a learning rate decay schedule, which reduces the learning rate by $20 \%$ every 20 epochs.

\textbf{Ablation Study:}
We use DIV2K \cite{agustsson2017ntire} and follow the standard procedure by extracting sub-images of $192\times192$ for training.
We iterate for 40 epochs over the training dataset.
Since we compare with DWSR, we use $L1$-loss as the learning objective, as reported by the authors of DWSR. 

\textbf{DWSR-Scenario:}
We use DIV2K \cite{agustsson2017ntire} like in the ablation study, but we train for 100 epochs as reported in DWSR.

\textbf{MWCNN-Scenario:}
We collect 800 images from DIV2K \cite{agustsson2017ntire}, 200 images from BSD \cite{martin2001database} and 4,744 images from WED \cite{ma2016waterloo} and train for 100 epochs.
Contrary to DWSR, we adapt the $L2$-loss like the authors of MWCNN.
For sub-image extraction, we use a size of $240\times240$ to match the training settings of MWCNN.
%
%
%
%

\begin{table}[!t]
\begin{center}
    \caption{\label{tab:ablStriding} Comparison of different striding settings on BSDS100 (2x and 4x scaling).}
    \begin{tabular}{c  l  c c c c  }
    \toprule
    Scale & Metric & no stride & s=1 & s=2 & s=3\\
    \midrule
    \multirow{2}{*}{2x} & \textbf{PSNR} $\uparrow$ & 31.8314 & \textbf{31.8660} & 31.8598 & 31.8588\\
    & \textbf{SSIM} $\uparrow$ & 0.9058 & \textbf{0.9061} & 0.9060 & 0.9059\\
    \midrule
    \multirow{2}{*}{4x} & \textbf{PSNR} $\uparrow$ & 27.2870 & \textbf{27.3048} & 27.2927 & 27.2872\\
    & \textbf{SSIM} $\uparrow$ & 0.7457 & \textbf{0.7471} & 0.7464 & 0.7466\\
    \bottomrule
    \end{tabular}
\end{center}
\end{table}

\section{Results}
This section presents the quantitative and qualitative analysis of this work. It shows that incorporating the DWA module into DWSR improves the performance in every dataset and for all scaling factors. 
Moreover, we consistently improve the SSIM scores by implementing DWA into MWCNN and achieve similar PSNR results. 
This section starts with an ablation study to investigate different striding settings and the effect of combining DWA with DWSR for the direct application and the regular DWT case (see \autoref{sec:direct}).
Next, we examine the performance scores of our DWA module on classical SR datasets with DWSR and MWCNN.
Finally, we visually compare the quality of the reconstructions.

\subsubsection{Ablation Study}
\autoref{tab:ablStriding} shows the impact of different striding settings for DWSR with DWA with 2x and 4x scaling.
We observe an improvement for striding settings greater than 0, significantly for PSNR and slightly for SSIM. 
The differences between striding settings greater than 0 are minimal, with a slight decrease for larger striding sizes.
Nonetheless, they outperform DWA with no stride difference consistently.
Thus, having a stride difference to capture local variations more effectively benefits the overall performance of DWSR.

\begin{figure}
    \begin{center}
        \includegraphics[width=.95\textwidth]{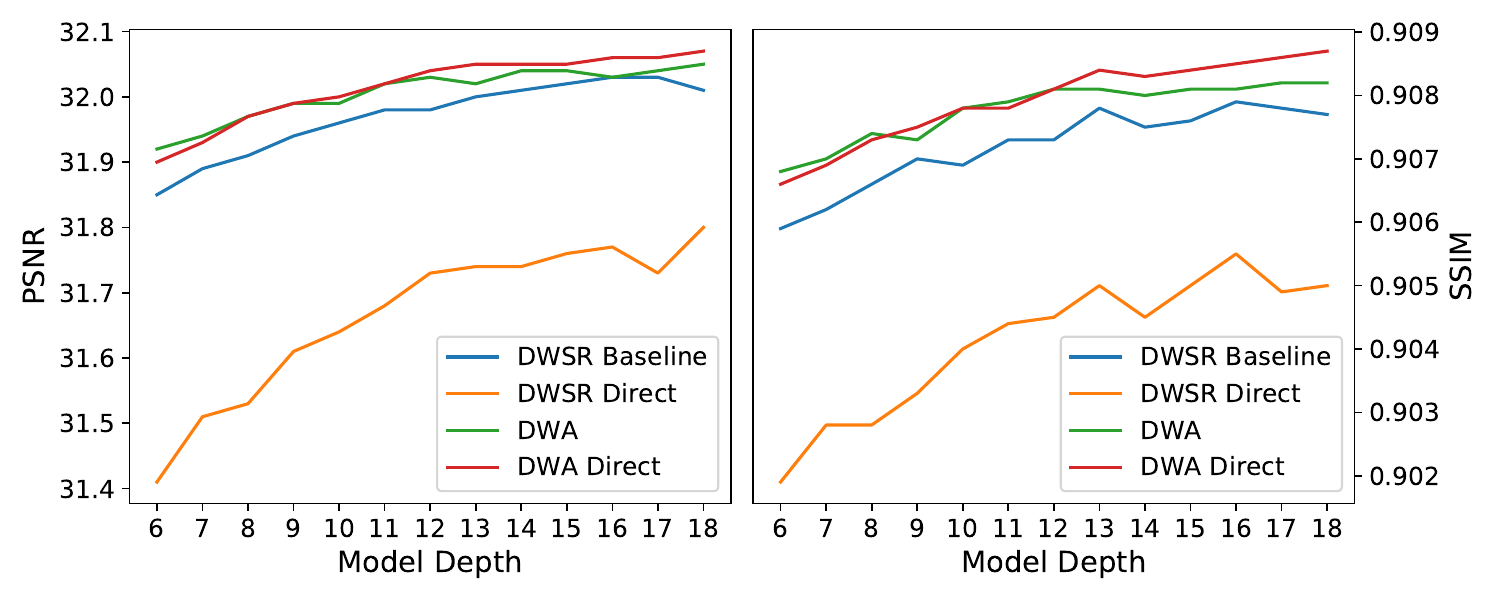}
        \caption{\label{fig:ablation}
        Results of ablation study on BSDS100 with scaling factor 2x. We tested different configurations: Baseline, Direct (application on the image space), DWA, and DWA Direct (application on the image space). 
        }
    \end{center}
\end{figure}

We further investigate the impact of various model configurations, DWSR with or without the DWA module, in a direct application or without (see \autoref{sec:direct}).
\autoref{fig:ablation} presents the results, where two graphs display the PSNR and SSIM values \cite{10041995}, respectively, for each method.
We apply the ablation study with different model depths, ranging from 6 to 18, instead of using a standard depth of 10 for DWSR.
As a result, DWSR with DWA or DWA Direct consistently outperforms the DWSR baseline across all model depths. 
This demonstrates the effectiveness of incorporating the DWA module as the first layer in the DWSR framework. 
Moreover, DWA Direct outperforms DWA applied to the DWT on the input with greater model depths.
Furthermore, we observe a considerable performance drop in DWSR Direct without using the DWA module compared to all other evaluated methods. 
This indicates that the DWA module is crucial in enabling the Direct approach, as its absence significantly degrades performance.
\subsubsection{Performance}
\autoref{tab:perf} summarizes PSNR and SSIM scores when applying the DWA module to DWSR and MWCNN for classical SR datasets on different scaling factors for a longer training span. 
We observe that incorporating the DWA module into DWSR improves the performance in every dataset and for all scaling factors.
For MWCNN with DWA, a similar observation can be made, especially for the SSIM scores, which show overall the best performances. 
However, it has slightly decreased PSNR values for some cases, e.g., for scaling factor 3.
Both applications, DWSR with DWA and MWCNN with DWA, are applied directly on the input image space, omitting a DWT of the input.

\begin{table*}
\begin{center}
    \caption{\label{tab:perf}Comparison of DWSR, MWCNN, and DWA Direct with DWSR and MWCNN architecture on Set5, Set14, and BSDS100.
    Note that PSNR [dB] is a logarithmic scale, and SSIM reflects correlations (with values ranging from -1 to 1) \cite{10041995}.
    }
    \begin{tabular}{ l | c | c |  c | c | c  }
    \toprule
    \multirow{3}{*}{Dataset} & \multirow{3}{*}{Scale}  & DWSR & MWCNN & DWA Direct & DWA Direct \\
    & & \cite{guo2017deep} & \cite{liu2018multi} & [DWSR] & [MWCNN] \\
    & & PSNR/SSIM & PSNR/SSIM & PSNR/SSIM & PSNR/SSIM \\
    \midrule
    \multirow{3}{*}{Set5 \cite{bevilacqua2012low}} 
    & 2x & 37.43 / 0.9568 & 37.91 / 0.9600 & 37.79 / 0.9645 & \textbf{37.99} / \textbf{0.9652} \\
    & 3x & 33.82 / 0.9215 & \textbf{34.18} / 0.9272 & 33.85 / 0.9310 & 34.09 / \textbf{0.9329} \\
    & 4x & 31.39 / 0.8833 & 32.12 / 0.8941 &  31.76 / 0.8898 & \textbf{32.16} / \textbf{0.9054} \\
    \midrule
    \multirow{3}{*}{Set14 \cite{zeyde2010single}} 
    & 2x & 33.07 / 0.9106 & \textbf{33.70} / 0.9182 & 33.38 / 0.9237  & \textbf{33.70} / \textbf{0.9265} \\
    & 3x & 29.83 / 0.8308 & \textbf{30.16} / 0.8414 & 29.90 / 0.8504 & 30.12 / \textbf{0.8545} \\
    & 4x & 28.04 / 0.7669 & 28.41 / 0.7816 & 28.31 / 0.7928 & \textbf{28.70} / \textbf{0.8012} \\
    \midrule
    \multirow{3}{*}{BSDS100 \cite{martin2001database}} 
    & 2x & 31.80 / 0.8940 & \textbf{32.23} / 0.8999 & 32.01 / 0.9080  & 32.21 / \textbf{0.9102} \\
    & 3x & n.a. & \textbf{29.12} / 0.8060 & 28.79 / 0.8174  & 28.93 / \textbf{0.8211} \\
    & 4x & 27.25 / 0.7240 & 27.62 / 0.7355 & 27.38 / 0.7503  & \textbf{27.63} / \textbf{0.7573} \\
    \bottomrule
    \end{tabular}
\end{center}
\end{table*}

\subsubsection{Visual Comparison}
\begin{figure}[hbt!]
    \begin{center}
        \includegraphics[width=.95\textwidth]{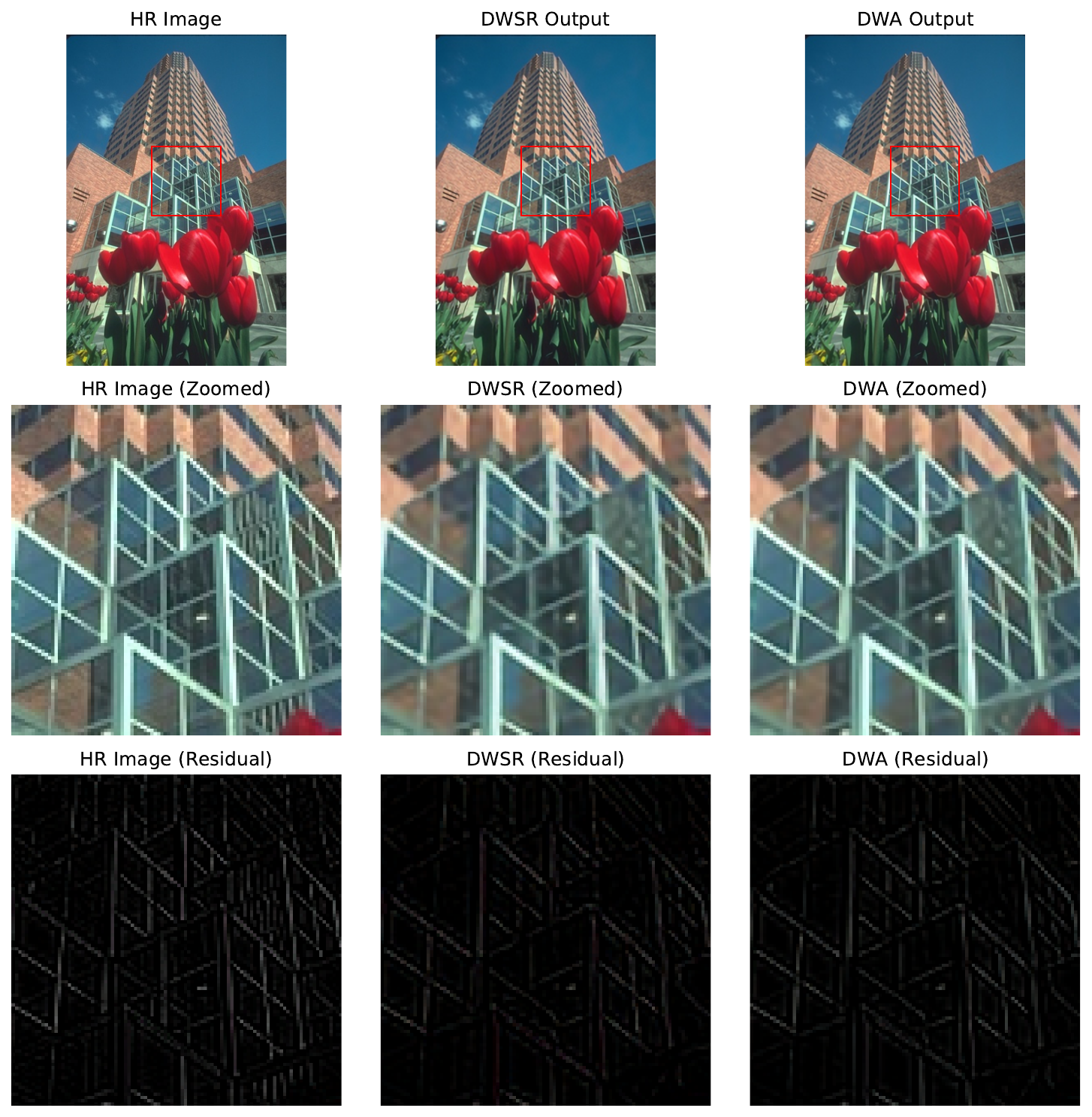}
        \caption{\label{fig:qual1}
        Comparison of an HR ground truth image (BSDS100, 2x scaling), DWSR, and DWA. First row: the entire image space of the HR image and the corresponding reconstructions obtained by the SR models. Second row: zoomed-in regions within the images from the first row. Third row: residual image representing the difference between the LR and HR images. As a result, the DWA model captures edges and details closer to the ground truth residuals, as opposed to the DWSR model (also regarding color).}
    \end{center}
\end{figure}
\autoref{fig:qual1} displays the ground truth HR image alongside the DWSR and DWA reconstructions. 
DWSR and DWA perform reasonably well in reconstructing the images. 
However, the DWA reconstructions exhibit more accurate and sharp details, particularly in the zoomed-in regions.
Since the added bicubic interpolation of the LR image in the reconstruction process provides a robust base prediction, we also present the residual images, which are the differences between the bicubic interpolations and the ground truth images, to highlight the performance difference between both approaches. 

\begin{figure}[hbt!]
    \begin{center}
        \includegraphics[width=.85\textwidth]{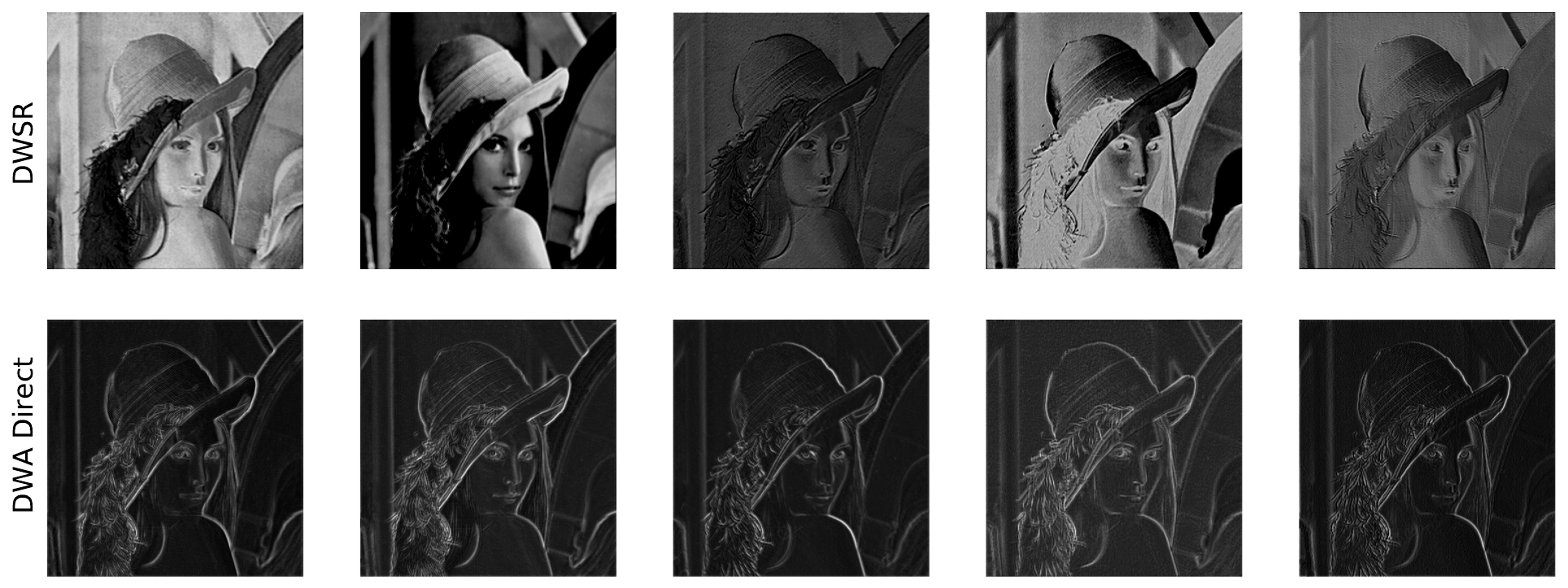}
        \caption{\label{fig:feats}
        Feature maps of DWSR and DWA Direct. 
        DWA Direct extracts local contrasts and variations more effectively, closer than DWSR to the residual target.
        }
    \end{center}
\end{figure}

These residual images are the learning targets of the models to improve the reconstruction quality beyond interpolation.
By comparing the residual images, we can see more clearly that the DWA model captures better distinct edges and finer details, which are also closer to the ground truth residuals, as opposed to the DWSR model.
It has more substantial edges and finer points in the residual images, which are also closer in color (see red colored lines of DWSR reconstruction in \autoref{fig:qual1} as a comparison).
This observation aligns with our quantitative results, where DWA outperforms DWSR regarding various performance metrics.

To provide deeper insights into our proposed models, \autoref{fig:feats} presents feature maps generated by the DWSR and DWA Direct models after the first layer. 
To ensure diversity, we selected the top five channels from each method based on the highest sum of distances between pairwise differences of all channels.
Our analysis reveals that although DWSR operates on the frequency space, it still remains similar to the LR input and fails to capture the desired target residual. 
In contrast, DWA Direct extracts local contrasts and variations more effectively from the image space and performs better in mapping the target residual. 
%
%
%
\section{Conclusion and Future Work}
In this work, we presented a novel Differential Wavelet Amplifier (DWA) module, which can be used as a drop-in module to existing wavelet-based SR models. 
We showed experimentally on Set5, Set14, and BSDS100 for scaling factors 2, 3, and 4 that it improves the reconstruction quality of the SR models DWSR and MWCNN while enabling an application of them to the input image space directly without harm to performance.
This module captures more distinct edges and finer details, which are closer to the ground truth residuals, which wavelet-based SR models usually learn.
This work is an opportunity to seek further advancements for SR based on frequency-based representations.

For future work, an exciting research avenue would be to explore ways to incorporate DWA on different DWT levels in MWCNN instead of only applying it initially.

\section*{Acknowledgments}
This work was supported by the BMBF projects SustainML (Grant 101070408) and by Carl Zeiss Foundation through the Sustainable Embedded AI project (P2021-02-009).

%
%
%
\bibliographystyle{splncs04}
\bibliography{mybibliography}

\begin{thebibliography}{10}
\providecommand{\url}[1]{\texttt{#1}}
\providecommand{\urlprefix}{URL }
\providecommand{\doi}[1]{https://doi.org/#1}

\bibitem{agarap2018deep}
Agarap, A.F.: Deep learning using rectified linear units (relu). arXiv preprint
  arXiv:1803.08375  (2018)

\bibitem{agarwal2005foundations}
Agarwal, A., Lang, J.: Foundations of analog and digital electronic circuits.
  Elsevier (2005)

\bibitem{agustsson2017ntire}
Agustsson, E., Timofte, R.: Ntire 2017 challenge on single image
  super-resolution: Dataset and study. In: Proceedings of the IEEE conference
  on computer vision and pattern recognition workshops. pp. 126--135 (2017)

\bibitem{bevilacqua2012low}
Bevilacqua, M., Roumy, A., Guillemot, C., Alberi-Morel, M.L.: Low-complexity
  single-image super-resolution based on nonnegative neighbor embedding  (2012)

\bibitem{canh2019difference}
Canh, T.N., Jeon, B.: Difference of convolution for deep compressive sensing.
  In: 2019 IEEE International Conference on Image Processing (ICIP). pp.
  2105--2109. IEEE (2019)

\bibitem{dong2015image}
Dong, C., Loy, C.C., He, K., Tang, X.: Image super-resolution using deep
  convolutional networks. IEEE transactions on pattern analysis and machine
  intelligence  \textbf{38}(2),  295--307 (2015)

\bibitem{dong2016accelerating}
Dong, C., Loy, C.C., Tang, X.: Accelerating the super-resolution convolutional
  neural network. In: European conference on computer vision. pp. 391--407.
  Springer (2016)

\bibitem{graves2008offline}
Graves, A., Schmidhuber, J.: Offline handwriting recognition with
  multidimensional recurrent neural networks. Advances in neural information
  processing systems  \textbf{21} (2008)

\bibitem{guo2017deep}
Guo, T., Seyed~Mousavi, H., Huu~Vu, T., Monga, V.: Deep wavelet prediction for
  image super-resolution. In: Proceedings of the IEEE Conference on Computer
  Vision and Pattern Recognition Workshops. pp. 104--113 (2017)

\bibitem{he2016deep}
He, K., Zhang, X., Ren, S., Sun, J.: Deep residual learning for image
  recognition. In: Proceedings of the IEEE conference on computer vision and
  pattern recognition. pp. 770--778 (2016)

\bibitem{horowitz1989art}
Horowitz, P., Hill, W., Robinson, I.: The art of electronics, vol.~2. Cambridge
  university press Cambridge (1989)

\bibitem{kingma2014adam}
Kingma, D.P., Ba, J.: Adam: A method for stochastic optimization. arXiv
  preprint arXiv:1412.6980  (2014)

\bibitem{knutsson1993normalized}
Knutsson, H., Westin, C.F.: Normalized and differential convolution. In:
  Proceedings of IEEE Conference on Computer Vision and Pattern Recognition.
  pp. 515--523. IEEE (1993)

\bibitem{laplante2018comprehensive}
Laplante, P.A.: Comprehensive dictionary of electrical engineering. CRC Press
  (2018)

\bibitem{li2022srdiff}
Li, H., Yang, Y., Chang, M., Chen, S., Feng, H., Xu, Z., Li, Q., Chen, Y.:
  Srdiff: Single image super-resolution with diffusion probabilistic models.
  Neurocomputing  \textbf{479},  47--59 (2022)

\bibitem{liang2021swinir}
Liang, J., Cao, J., Sun, G., Zhang, K., Van~Gool, L., Timofte, R.: Swinir:
  Image restoration using swin transformer. In: Proceedings of the IEEE/CVF
  international conference on computer vision. pp. 1833--1844 (2021)

\bibitem{liu2018multi}
Liu, P., Zhang, H., Zhang, K., Lin, L., Zuo, W.: Multi-level wavelet-cnn for
  image restoration. In: Proceedings of the IEEE conference on computer vision
  and pattern recognition workshops. pp. 773--782 (2018)

\bibitem{lowe2004distinctive}
Lowe, D.G.: Distinctive image features from scale-invariant keypoints.
  International journal of computer vision  \textbf{60},  91--110 (2004)

\bibitem{ma2016waterloo}
Ma, K., Duanmu, Z., Wu, Q., Wang, Z., Yong, H., Li, H., Zhang, L.: Waterloo
  exploration database: New challenges for image quality assessment models.
  IEEE Transactions on Image Processing  \textbf{26}(2),  1004--1016 (2016)

\bibitem{martin2001database}
Martin, D., Fowlkes, C., Tal, D., Malik, J.: A database of human segmented
  natural images and its application to evaluating segmentation algorithms and
  measuring ecological statistics. In: Proceedings Eighth IEEE International
  Conference on Computer Vision. ICCV 2001. vol.~2, pp. 416--423. IEEE (2001)

\bibitem{10041995}
Moser, B.B., Raue, F., Frolov, S., Palacio, S., Hees, J., Dengel, A.:
  Hitchhiker's guide to super-resolution: Introduction and recent advances.
  IEEE Transactions on Pattern Analysis and Machine Intelligence pp. 1--21
  (2023). \doi{10.1109/TPAMI.2023.3243794}

\bibitem{ronneberger2015u}
Ronneberger, O., Fischer, P., Brox, T.: U-net: Convolutional networks for
  biomedical image segmentation. In: Medical Image Computing and
  Computer-Assisted Intervention--MICCAI 2015: 18th International Conference,
  Munich, Germany, October 5-9, 2015, Proceedings, Part III 18. pp. 234--241.
  Springer (2015)

\bibitem{sahak2023denoising}
Sahak, H., Watson, D., Saharia, C., Fleet, D.: Denoising diffusion
  probabilistic models for robust image super-resolution in the wild. arXiv
  preprint arXiv:2302.07864  (2023)

\bibitem{saharia2022image}
Saharia, C., Ho, J., Chan, W., Salimans, T., Fleet, D.J., Norouzi, M.: Image
  super-resolution via iterative refinement. IEEE Transactions on Pattern
  Analysis and Machine Intelligence  (2022)

\bibitem{sarigul2019differential}
Sar{\i}g{\"u}l, M., Ozyildirim, B.M., Avci, M.: Differential convolutional
  neural network. Neural Networks  \textbf{116},  279--287 (2019)

\bibitem{stephane1999wavelet}
Stephane, M.: A wavelet tour of signal processing (1999)

\bibitem{sun2023spatially}
Sun, L., Dong, J., Tang, J., Pan, J.: Spatially-adaptive feature modulation for
  efficient image super-resolution. arXiv preprint arXiv:2302.13800  (2023)

\bibitem{vaswani2017attention}
Vaswani, A., Shazeer, N., Parmar, N., Uszkoreit, J., Jones, L., Gomez, A.N.,
  Kaiser, {\L}., Polosukhin, I.: Attention is all you need. Advances in neural
  information processing systems  \textbf{30} (2017)

\bibitem{Wang_2021_ICCV}
Wang, X., Xie, L., Dong, C., Shan, Y.: Real-esrgan: Training real-world blind
  super-resolution with pure synthetic data. In: Proceedings of the IEEE/CVF
  International Conference on Computer Vision (ICCV) Workshops. pp. 1905--1914
  (October 2021)

\bibitem{wang2018esrgan}
Wang, X., Yu, K., Wu, S., Gu, J., Liu, Y., Dong, C., Qiao, Y., Change~Loy, C.:
  Esrgan: Enhanced super-resolution generative adversarial networks. In:
  Proceedings of the European conference on computer vision (ECCV) workshops.
  pp.~0--0 (2018)

\bibitem{9044873}
Wang, Z., Chen, J., Hoi, S.C.H.: Deep learning for image super-resolution: A
  survey. IEEE Transactions on Pattern Analysis and Machine Intelligence
  \textbf{43}(10),  3365--3387 (2021). \doi{10.1109/TPAMI.2020.2982166}

\bibitem{zeyde2010single}
Zeyde, R., Elad, M., Protter, M.: On single image scale-up using
  sparse-representations. In: International conference on curves and surfaces.
  pp. 711--730. Springer (2010)

\end{thebibliography}
%
%
%
%
%
\end{document}